\documentclass[prl,aps,twocolumn,showpacs]{revtex4}
\usepackage{graphicx}
\usepackage{amsmath}
\usepackage{amssymb}
\usepackage{bm}
\newcommand{\be}{\begin{equation}}
\newcommand{\ee}{\end{equation}}
\newcommand{\bea}{\begin{eqnarray}}
\newcommand{\eea}{\end{eqnarray}}
\newcommand{\bei}{\begin{itemize}}
\newcommand{\eei}{\end{itemize}}

\begin{document}

\title{Signatures of Fractional Exclusion Statistics in the Spectroscopy of Quantum Hall Droplets}

\author{Nigel R. Cooper}
\affiliation{T.C.M. Group, Cavendish Laboratory, University of Cambridge, J.~J.~Thomson Ave., Cambridge CB3~0HE, United Kingdom}
\author{Steven H. Simon}
\affiliation{Rudolf Peierls Centre for Theoretical Physics, University of Oxford, 1 Keble Road, Oxford OX1~3NP, United Kingdom}

\begin{abstract}

We show how spectroscopic experiments on a small Laughlin droplet of rotating bosons can directly demonstrate Haldane fractional exclusion statistics of quasihole excitations.    The characteristic signatures appear in the single-particle excitation spectrum.  We show that the transitions are governed by a ``many-body selection rule'' which allows one to relate the number of allowed transitions to the number of quasihole states on a finite geometry. We illustrate the theory with numerically exact simulations of small numbers of particles.

\end{abstract}
\date{\today}
 
%67.85.-d       Ultracold gases, trapped gases
%73.43.Cd     (QH Effects) Theory and modeling
%05.30.Pr     Fractional statistics systems (anyons, etc.)

\maketitle

One of the most dramatic features of strongly correlated phases is the
emergence of quasiparticle excitations with unconventional quantum
statistics.  The archetypal example is the fractional, ``anyonic'',
quantum statistics predicted for the quasiparticles of the fractional
quantum Hall phases\cite{halperin,arovas}. While experiments on semiconductor devices have
shown that these quasiparticles have fractional charges\cite{shotnoise1,shotnoise2,amir}, a direct
observation of the fractional statistics has remained lacking.

In this Letter we show how precision spectroscopy measurements of
rotating droplets of ultracold atoms could be used to demonstrate the
Haldane fractional exclusion statistics\cite{Haldane} of quasiholes in the Laughlin state of
bosons.  By involving only spectroscopic signatures of the rotating
droplet, our proposal plays to the strengths of atomic physics
experiments. We show that evidence of the fractional exclusion
statistics appears in counting the numbers of lines in the
radio-frequency (RF) absorption spectrum. In this sense, the method is
conceptually similar to classic evidence of quantum statistics, as
appearing in the rotational levels of homonuclear diatomic molecules
(e.g. the Fermi statistics of the proton causing the rotational levels
of H$_2$ to depend on whether the spins of the nuclei are in singlet
or triplet state). Our method differs substantially from proposals to measure the fractional braiding statistics of quasiholes\cite{paredes,kapit,grass}, notably by not requiring local time-dependent potentials for the adiabatic manipulation of the positions of the quasiholes.

We have in mind a fast rotating gas of identical bosonic atoms,
initially in a single internal (hyperfine) state $\Uparrow$, and confined to a
quasi-2D layer with oscillator length $a_z$. The gas is subjected to a tight circularly symmetric harmonic trap of
frequency $\omega_0$, with $\hbar \omega_0 \gg V_0$, in which
$V_0\equiv \sqrt{\frac{2}{\pi}}\frac{a_{\rm s}}{a_z}\hbar\omega_0$ is a characteristic interaction energy 
 for atoms with scattering length $a_{\rm s}$. Hence, the
interactions leave the particles in the lowest Landau level (LLL)\cite{advances}. In
addition, we shall consider a weak quartic potential ---
weak compared to both $\hbar\omega_0$ and $V_0$ --- for reasons to
be described below.  Specifically, we shall consider an initial state
of $N_{\rm i}$ atoms which has been spun up to the angular momentum
$L_{\rm i}=N_{\rm i}(N_{\rm i}-1)$. Then, for the case of contact repulsive interactions relevant in typical cold gas experiments, the groundstate is the (exact) $\nu=1/2$ Laughlin state.
Furthermore, for the case of contact interactions, the quasihole
excitations of these Laughlin droplets are non-interacting: this will
allow us to find evidence of the fractional exclusion statistics even
in small systems of $N_{\rm i} \lesssim 10$ atoms.
Experimental protocols to generate this initial state for small
numbers of atoms have been identified\cite{popp:053612,baur-2008},
and experimental work on driven lattices\cite{gemelke} has
investigated the properties of rapid rotation on multi-droplet
systems.  We shall focus on the properties of a single droplet, and
the spectroscopic signatures we seek shall require single-atom
imaging; such conditions are likely to require technologies developed
in ultracold gas microscopes\cite{Bakr09,Sherson10}.

\begin{figure}
\includegraphics[width=0.8\columnwidth]{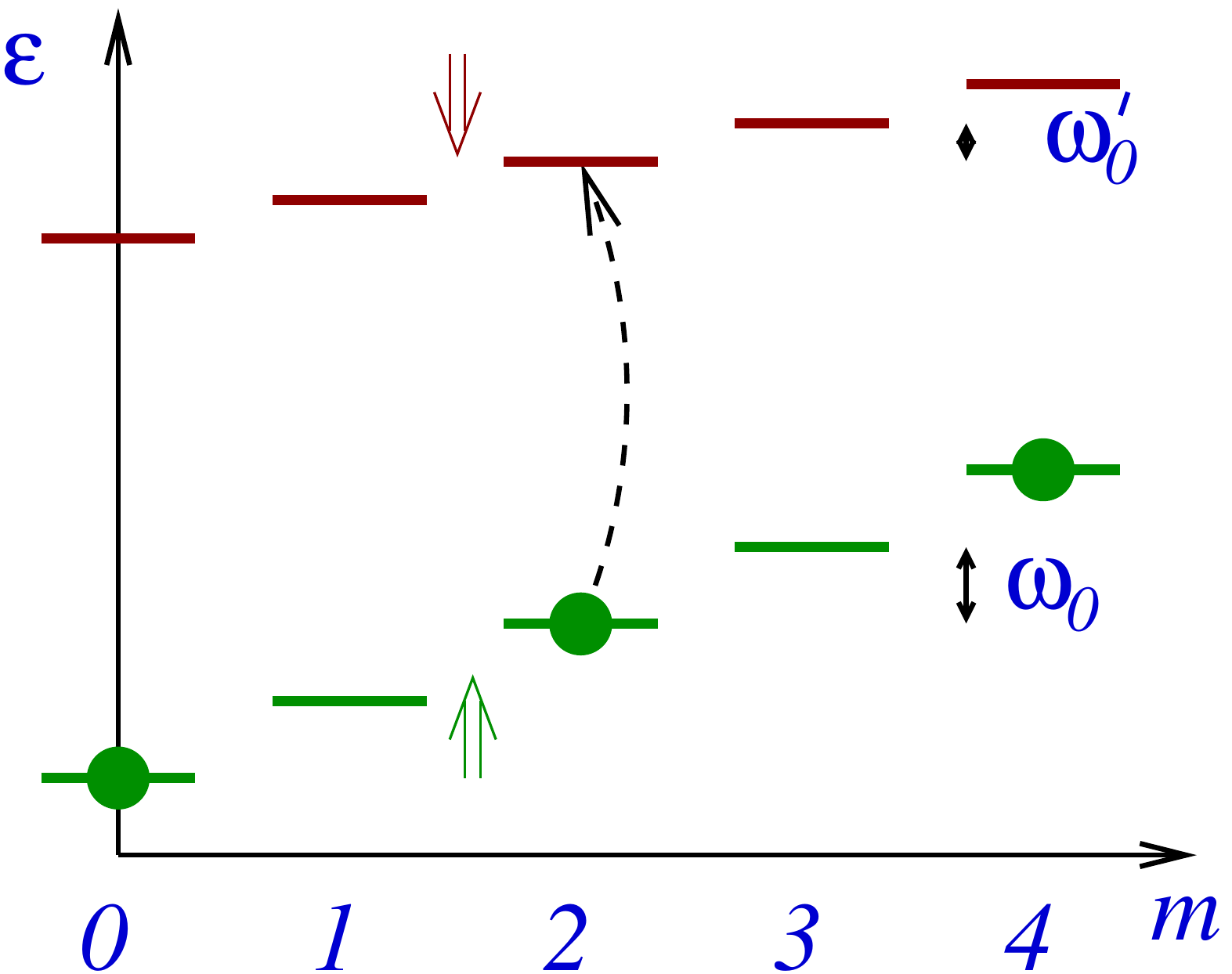}
\caption{ \label{fig:schematic-spectrum} Schematic illustration of RF excitation of a single particle
 from state $\Uparrow$ to $\Downarrow$.  The energy of the transition reveals the angular momentum of the excited particle.}
\label{fig:spect}
\end{figure}

Now consider making an RF excitation of a single atom from internal
state $\Uparrow$ into an internal state $\Downarrow$ which does not
interact with the initial $\Uparrow$ atoms. For hyperfine states, such
situations can be found by tuning to the zero of a Feshbach resonance.
The promoted atom can carry away an angular momentum, $m_{\rm f}$, in
the range $0\leq m_{\rm f} \leq m_{\rm max}$,
leaving the $N_{\rm f} =
N_{\rm i}-1$ atoms with angular momentum $L_{\rm f} = L_{\rm i} -
m_{\rm f}$ in the range $L_{\rm i} - m_{\rm max} \leq L_{\rm f} \leq L_{\rm i}$.  (The upper limit $m_{\rm max} = 2(N_{\rm i}-1)=  2N_{\rm f}$ is the highest angular momentum carried by any one particle in
 the initial Laughlin state of $N_{\rm i}$ particles\cite{Note1}.)
In the
following, we shall imagine that the transition spectrum can be
resolved into components labelled by this final angular momentum
$L_{\rm f}$. In principle this could be done by measuring the final
angular momentum of the excited $\Downarrow$ atom. However, note that,
in general, the change in internal states of the atom will also change
the confinement frequency. If the new confinement frequency
$\omega_0'$ is such that $\hbar|\omega_0-\omega_0'|\gg V_0$, then
the final angular momentum of the excited particle can be found
spectroscopically. This is illustrated in Fig.~\ref{fig:spect}, which
plots the single particle energies $\epsilon_{\Uparrow, m} =
\epsilon_{\Uparrow,0} + m\hbar\omega_0$ and $\epsilon_{\Downarrow, m}
= \epsilon_{\Downarrow,0} + m\hbar\omega_0'$, and indicates an RF
transition, for which there is no change in orbital angular momentum
$\delta m =0$.   Note that we also should assume  $|\omega_0-\omega_0'| \ll \omega_0$ so that the $|\Uparrow,m\rangle$ and the $|\Downarrow,m\rangle$ orbitals are roughly aligned spatially.

In considering the form of the spectrum of these RF transitions there
are two questions of importance: what are the energies of the final
states; and what are the matrix elements for transitions into these
final states?

Among the possible final states, there is a set with zero interaction
energy. These are of particular interest: they are the edge and quasihole
excitations of the Laughlin state of $N_{\rm f}$ particles, and their
properties provide a robust characterization of the Laughlin state.
For $L_{\rm f} = N_{\rm f}(N_{\rm f}-1)$ the only possible final state with zero interaction
energy is the Laughlin state for $N_{\rm f}$ particles. As $L_{\rm f}$ increases, the
number of zero interaction energy states increases, with well-defined
counting rules, which for $N_{\rm f}=5$ lead to the multiplicities shown in
Table~\ref{tab:states}.  This integer sequence is highly indicative of
the Laughlin state, giving information on the edge/quasihole
excitations\cite{Note2}.
Note that an analogous experiment performed with a Laughlin state at
a different filling factor $\nu=1/p$ would show the
same series of multiplicities (for the same $N_{\rm f}$) although the final angular momentum will range from $p N_{\rm f}(N_{\rm f}-1)/2$ to $pN_{\rm{f}}(N_{\rm{f}}+1)/2$, covering the first  $pN_{\rm{f}}+1=m_{\rm{max}}+1$ values of this sequence. 

\begin{figure}[htbp]
\includegraphics[width=0.8\columnwidth]{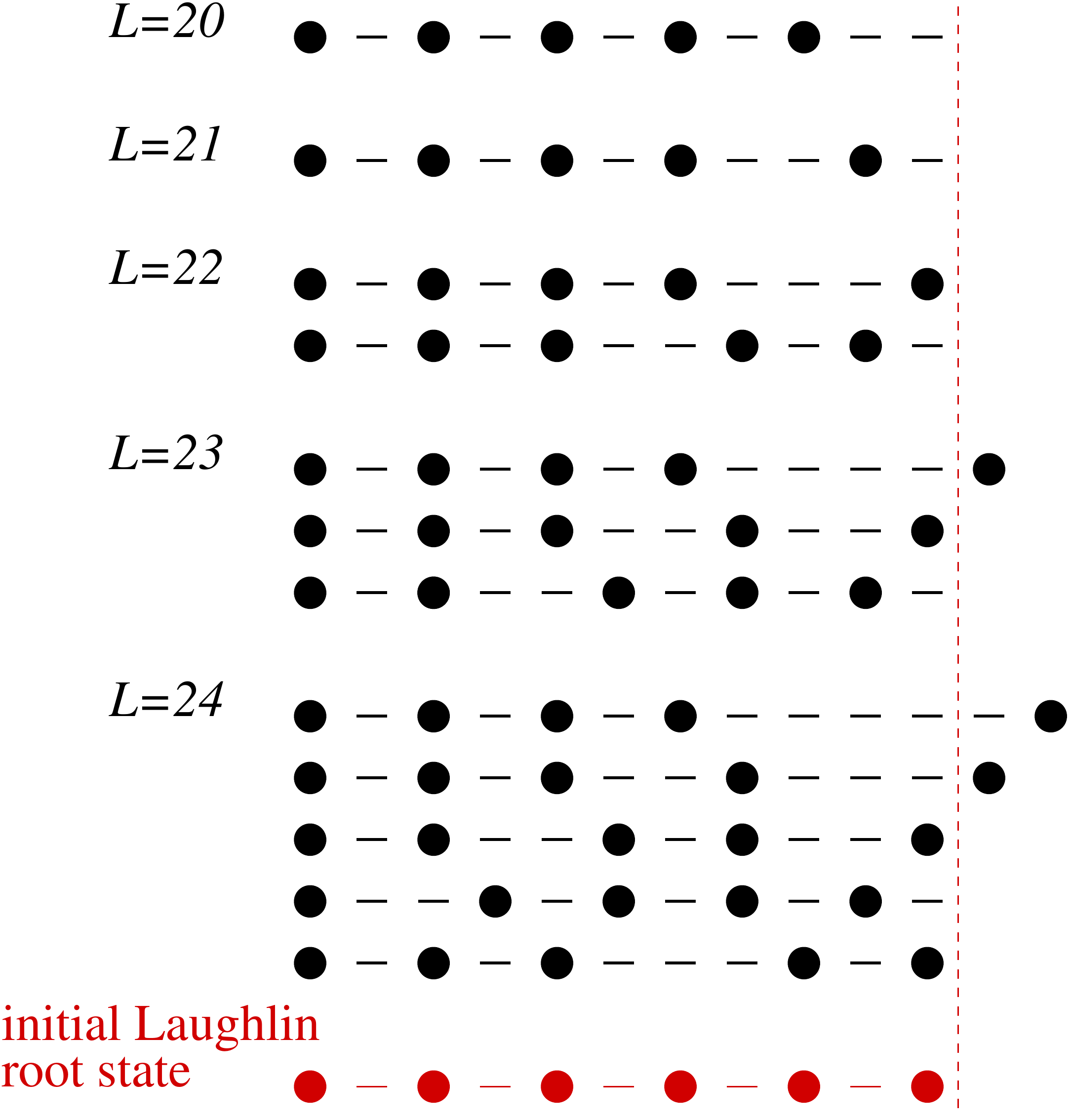}
\caption{ \label{fig:count5} Counting of the states for $N_{\rm f}=5$ contact interacting bosons for different final angular momenta $L$. Shown in red is the root state for the initial Laughlin state of $N_{\rm i}=6$ particles. This sets the cut-off (red dashed line) for the root configurations with sizeable matrix element: the allowed final states are those for which all particles are to the left of the red dashed line.}
\end{figure}

The counting of the zero interaction energy states can be
obtained from a simple picture based on the generalized clustering
principle\cite{haldanecluster}, which identifies the ``root states''
of the exact quantum states. These root states are single Fock states,
with definite particle number $n_m$ in orbitals $m=0,1,2\ldots$.  For
the $\nu=1/2$ Laughlin state, the clustering principle is that no two
particles can be in orbitals, $m$ and $m'$, with $|m-m'|<2$. The
resulting root configurations for $N_{\rm f}=5$ are shown in
Fig.~\ref{fig:count5} for small total angular momentum.    Note that (nonorthogonal) basis vectors of the zero interaction energy space
can be made from the root Fock state superposed with daughter Fock states that do not generally obey the clustering principle.  
 The  daughter states have the interesting feature that they are always ``squeezed" from the root state\cite{haldanecluster}, meaning that if we write out a string to represent the occupancies of single particle orbitals $m$, squeezing always numerically reduces the value of this string while preserving the total number of particles as well as the total angular momentum, $L$.  For example if we write the root state string $1\, 0 \, 1 \, 0 \,0 \,1$ to mean we have filled the orbitals $m=5, m=3, m=0$ each once, a daughter state squeezed from this would be the string $ 1\, 0\,  0\, 1 \, 1\, 0$ which is numerically less than $1\, 0 \,1 \,0 \,0 \,1$.

\begin{table}[ht]
% title of Table
\centering
% used for centering table
\begin{tabular}{c | c c c}
% centered columns (4 columns)
\hline\hline
%inserts double horizontal lines
{$L_{\rm f}$} & {$\#_{N_{\rm f},L_{\rm f}}$} &{$\#^{\rm allowed}_{N_{\rm f},L_{\rm f}}$} & $L^{\rm s}_z=L_{\rm f}-N_{\rm f}^2$ \\ [0.5ex]
% inserts table                                                                                                         
%heading                                                                                                                  
\hline
% inserts single horizontal line                                                                                          
{20} &{1} & {1}  &-5  \\
{21} & {1}& {1}  &-4  \\
{22} & {2}& {2}  &-3  \\
{23} & {3}& {2}  &-2  \\
{24} & {5}& {3}  &-1  \\
{25} & {7}& {3}   &0  \\
{26} &{10}& {3}   &1  \\
{27} &{13}& {2}   &2 \\
{28} &{18}& {2}   &3 \\
{29} &{23}& {1}   &4\\
{30} & {30}& {1}  &5\\
{31} & {37} &{0} &  \\
{\vdots} &{\vdots}& {\vdots}\\
\hline
%inserts single line                                                                                                      
\end{tabular}
%\end{table}
\caption{The number of zero-energy final states of $N_{\rm f}=5$ contact interacting bosons $\#_{N_{\rm f}=5,L_{\rm f}}$,
and the number of these states with allowed transitions, $\#^{\rm allowed}_{N_{\rm f}=5,L_{\rm f}}$, following the many-body selection rule described in the text.  The final column gives the $z$-component of the angular momentum of the effective spherical system described in the text.   No states are allowed for $L_{\rm f} > 30$ and here there is no meaningful value for $L^{\rm s}_z$
}
\label{tab:states}
\end{table}

\begin{figure}
\includegraphics[width=0.9\columnwidth]{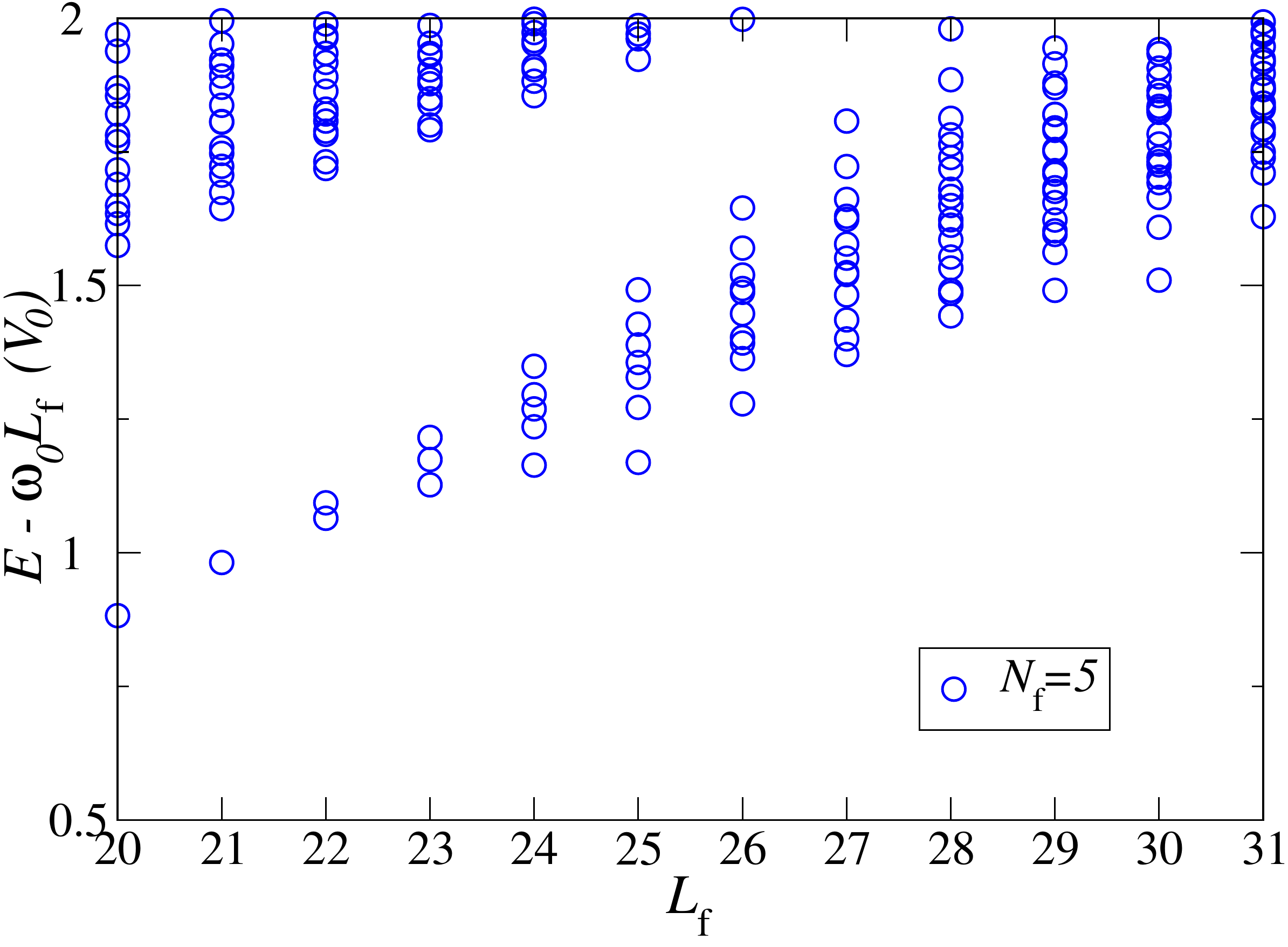}
\caption{ \label{fig:spectrum} Spectrum of final states of $N_{\rm f}=5$
  contact interacting bosons, showing the branch of zero interaction
  energy states below the (uninteresting) upper branch of high energy
  states. The zero interaction energy states (lower branch) are split
  in energy by a quartic potential. The $x$-axis is the final
  total angular momentum; the $y$-axis is energy in units of $V_0$.} \end{figure}

If there were no quartic potential, all of the final states with zero
interaction energy at a given $L_{\rm f}$ would be at exactly the same
energy (since interaction energy is zero). The quartic potential splits
these degenerate states, allowing separate transitions to be resolved.
Fig.~\ref{fig:spectrum} shows the splitting for $N_{\rm f}=5$. 
For weak quartic potential (compared to $V_0$) this splitting does not obscure
the many-body gap separating these zero-interaction energy states from the
high energy states.  Therefore, by detecting the number of low-energy spectral
lines (i.e. those below the gap, arising from many-body repulsion, in
Fig.~\ref{fig:spectrum}) as a function of $L_{\rm f}$, if all transitions had
non-zero matrix elements, one could measure the above counting sequence.

For a spectroscopic probe, it is important to determine the rate of transition into
the possible final states. This rate is proportional to the squared
matrix element
$${|\langle \mbox{final};{N_{\rm f}=N_{\rm i}-1,L_{\rm f}=L_{\rm i}-m_{\rm f}} | \hat{b}_{m_{\rm f}} |\mbox{Laughlin};{N_{\rm i}}\rangle|^2}$$
where $\hat{b}_m$ destroys a boson in state $m$.  We find that there
are very significant restrictions on such matrix elements. This leads to a strong ``many-body selection rule''
on the RF transitions in Laughlin clusters. Specifically, we find that
strong transitions --- which we shall refer to as ``allowed''
transitions --- exist only to those states whose root configurations
have all particles within the $m=0 \to 2N_{\rm f}$ orbitals.

The origin of the selection rule lies in the fact that the initial
Laughlin state, with $N_{\rm i}=N_{\rm f}+1$ particles, is a state in which all
particles are in these $m=0\ldots 2N_{\rm f}$ orbitals (see
Fig.~\ref{fig:count5}).  Hence, removal of a particle from the Laughlin state cannot
produce a Fock state with a particle in orbital $m> 2N_{\rm f}$.   Since we
find, from numerical diagonalizations, that the root configuration has
a very large weight in the exact eigenstates, matrix elements are
large only with those states whose root configurations have all
particles within $0 \leq m_{\rm f} \leq 2N_{\rm f}$\cite{Note3}.

The allowed states are a well-defined subset of the
zero-interaction energy states. This subset can be readily found from
Fig.~\ref{fig:count5} by retaining only those root states for which all
particles lie to the left of the red dashed line.  For this case of
$N_{\rm f}=5$ particles, the number of states that have allowed
transitions is shown in Table~\ref{tab:states}.

The set of allowed states has a simple interpretation: it is the
set of zero-energy states that can be formed for contact interacting
bosons on a system of fixed area with $2$ quasiholes (e.g. a sphere
with flux $N_\phi = 2 N_{\rm f}$). This follows from the fact that the number
of orbitals over which the $N_{\rm f}$ particles can be distributed is
$2N_{\rm f}+1$, while the (unique) Laughlin groundstate is formed when the number of
states is $2N_{\rm f}-1$. These $(2N_{\rm f}+1)-(2N_{\rm f}-1)=2$ excess orbitals can be viewed
as two quasiholes in the Laughlin groundstate. Each can be placed in
\begin{equation}
d_{\rm qh}=N_{\rm f}+1
\label{eqn:dqh}
\end{equation}
possible locations, with respect to the $N_{\rm f}$
particles. For two quasiparticles, the total number of states is simply given by
  the number of ways to put two identical quasiholes in $d_{\rm qh} =
  N_{\rm f}+1$ states: $\#^{\rm allowed}_{N_{\rm f}} \equiv \sum_{L_{\rm f}} \#^{\rm
    allowed}_{N_{\rm f},L_{\rm f}} = \frac{1}{2}(N_{\rm f}+1)(N_{\rm f}+2)$. This can be resolved into
  states of fixed angular momentum $L_{\rm f}$. Again we exploit  the  equivalence to the states of $N_{\rm f}$ particles on a sphere of
flux $N_\phi=2N_{\rm f}$: with $L^{\rm s}_z = -N_{\rm f} \ldots N_{\rm f}$ replacing $m=0 \ldots
2N_{\rm f}$, such that the $z$-projection of total angular momentum on the
sphere is $L^{\rm s}_z = L_{\rm f} - N_{\rm f}^2$. Since the Laughlin groundstate for $N_{\rm f}$
particles is at $N_\phi = 2(N_{\rm f}-1)$, the flux $N_\phi=2N_{\rm f}$ corresponds to
the addition of $n=2$ quasiholes.
  It is known\cite{ReadR96} that the total number
of zero energy states for $n$ quasiholes is given 
by the binomial coefficient $C(N_{\rm f}+n, n)$  (i.e., choose $n$ from $N_{\rm f}+n$). 
For $n=2$ these states can be indexed\cite{ReadR96} as states of total angular
momentum 
$L^{\rm s}_z=1,3,5,\ldots N_{\rm f}$ ($N_{\rm f}$ odd), or
$L^{\rm s}_z=0,2,4,\ldots N_{\rm f}$ ($N_{\rm f}$ even), which is consistent with
the counting in Table~\ref{tab:states}.

The relation (\ref{eqn:dqh}), together with the fact that for a system of fixed area, 
the removal of a single particle ($\Delta N=-1$) corresponds to the creation of two quasiholes $\Delta n_{\rm qh} = -2\Delta N$
fixes the Haldane exclusion statistics of the quasiholes. These generalized exclusion statistics  relate the change in dimension 
of the Hilbert space $\Delta d$ available to a quasiparticle to the change in number of quasiparticles 
$\Delta n$ via $\Delta d = -g \Delta n$ with $g$ being the exclusion statistic paramater.   For bosons or fermions, $g=0$ or $1$ respectively.  
Here we have $\Delta d = -\frac{1}{2} \Delta n$ showing that $g = \frac{1}{2}$, indicative of
``semionic'' statistics of the quasiholes.  Thus, by counting the number of allowed
transitions in the RF spectra, one obtains direct evidence for the
counting formula (\ref{eqn:dqh}). As described above, the dependence of
the counting formulas on $N_{\rm f}$ lies at the heart of the fractional
exclusion statistics for the quasholes: thus, by detecting the
multiplicities for different $N_{\rm f}$ (i.e.  different initial $N_{\rm i} =
N_{\rm f}+1$), amounts to a direct detection of these exclusion statistics.

\begin{figure}
\includegraphics[width=0.8\columnwidth]{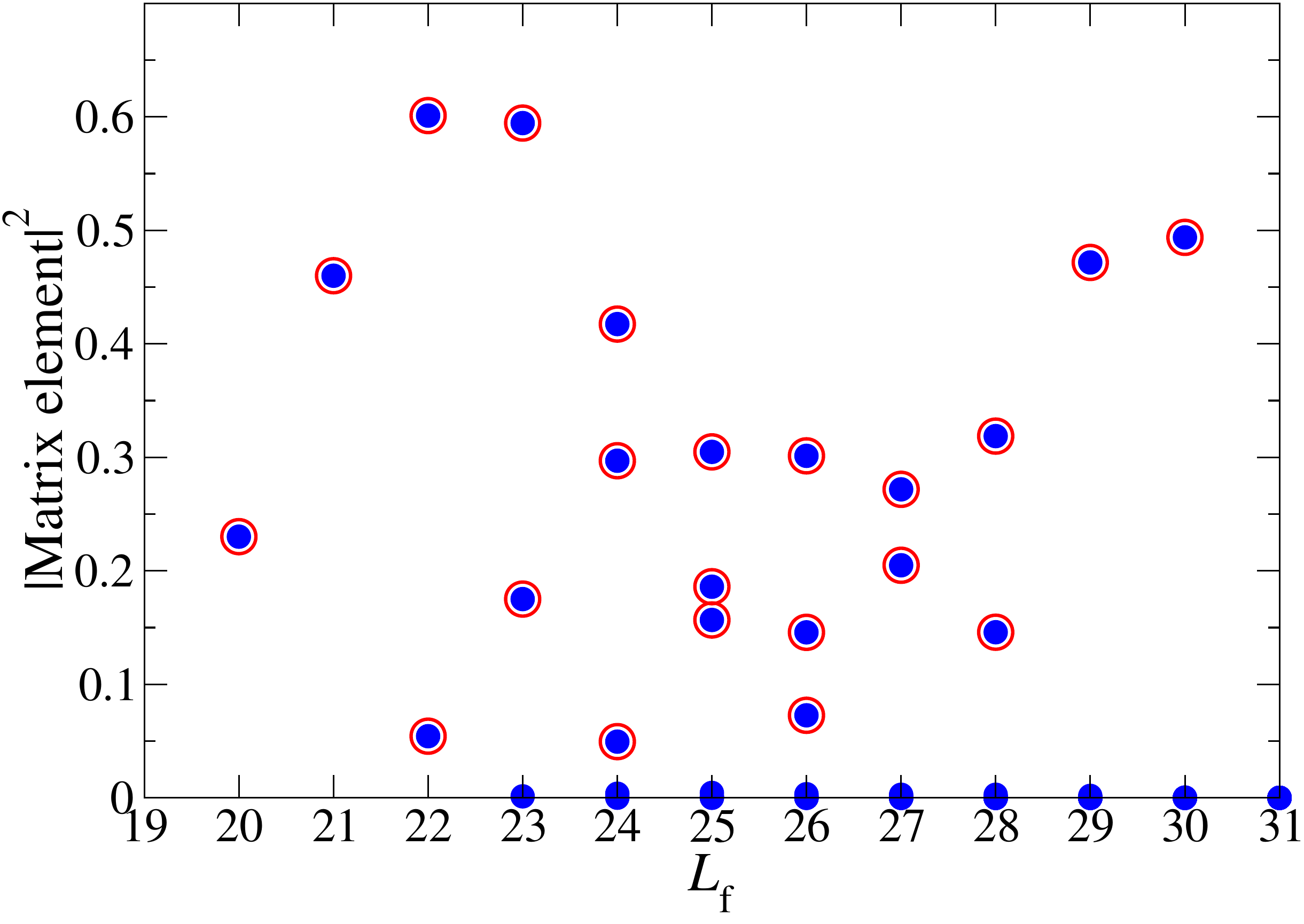}
\caption{ \label{fig:selection} Squared matrix element of transitions into the quasihole sector as a function of final angular momentum (blue dots) for $N_{\rm f} = 5$ particles in the final state.
The red circles indicate those transitions which are allowed according to the many-body selection rule described in the text.}
\end{figure}

The preceding discussion is based on the existence of the many-body
selection rule. How well does this many-body selection rule apply in
practice? To test this, we have computed the many-body states and
matrix elements numerically for $N_{\rm i} = 2\ldots 10$ particles.  In
Fig.~\ref{fig:selection} we show the matrix elements (squared) for all
excited states at each final momentum $L$, computed by exact
diagonalization for the case of $N_{\rm f}=5$. (Results for other values of
$N_{\rm f}$ are consistent.) There is a clear separation between allowed
states, with matrix element squared of order one, and forbidden
states, with square matrix element smaller by at least two orders of
magnitude, and imperceptible on the linear scale of
Fig.~\ref{fig:selection}. The counting of the allowed states (marked
by red circles in Fig.~\ref{fig:selection}) follows the pattern
(1,1,2,2,3,3,3,2,2,1,1) expected from the counting rules described
above (see Table~\ref{tab:states}).  The experimental goal will be to count the
transitions with nonnegligible weights in each angular momentum sector.

One might wonder why the many-body selection rule works as well as it does (with the matrix elements for states that violate the selection rule being suppresed by factors of 100 or more).   Our above argument that each eigenstate contains a large component of its root Fock state turns out not to be a sufficient explanation since there is substantial mixing with daughter states\cite{haldanecluster}.   Let us examine 
a more general potential $V \sim r^\gamma$ instead of the quartic potential $\gamma=4$ (while keeping the potential the smallest energy scale of the system).    In terms of orbital occupations $n_m$, this yields potential energies $\sim \sum_m m^{\gamma/2} \, n_m$.   In the limit  $\gamma \rightarrow \infty$ this is  dominated by the occupied orbitals with the largest $m$, and the potential energies of Fock states are then ordered with the same ``squeezing" relationship as described above.    Beginning with the basis of zero interaction energy states described by root states and their corresponding daughters, in the $\gamma\rightarrow \infty$ limit the {\it exact} energy eigenstates can then be found by successively orthogonalizing these wavefunctions, starting with the root state with lowest potential energy and continuing to successively higher root states.  Since removing a particle from the Laughlin state must generate a superposition of wavefunctions in this zero-interaction energy space which also 
contains no occupied orbitals with $m > N_{\rm f}$ the selection rule becomes exact.  We find numerically that this ``orthogonalized root state basis" is an extremely accurate representation of the exact eigenstates even when $\gamma=4$, providing some justification for why the selection rule works so well.

In summary, the spectroscopic probe removing one atom creates two quasiholes
in the Laughlin cluster. Measurements of the number of allowed transitions
would determine the number of these quasihole states, testing theories of the
properties (degeneracy and mutual statistics) of this pair of fractionalized
particles. While experiments of this kind are certainly technically
challenging --- requiring control of small numbers of atoms with high
fidelity, and the detection of single atoms in spectroscopic probes --- these
are within reach of new technologies of quantum gas microscopes.  Our work
provides a very appealing direct link between multiplicities in RF spectra,
and counting formulas for fractionalized excitations in strongly correlated
many-body systems.

\vskip0.1cm

\acknowledgments{We are grateful to Jean Dalibard for many enlightening comments. This work was supported by EPSRC Grants EP/J017639/1, EP/I032487/1,
EP/I031014/1, and by the Royal Society.}

\end{document}